# Ultralow power and shifting-discretized magnetic racetrack memory device driven by chirality switching and spin current

Shen Li[1,2,3], Xiaoyang Lin[1,3*], Pingzhi Li[2], Suteng Zhao[1], Zhizhong Si[1], Guodong Wei[1], Bert Koopmans[2], Reinoud Lavrijsen[2] and Weisheng Zhao[1,3]

[1] School of Integrated Circuit Science and Engineering, Beihang University, Beijing, 100191, China

[2] Department of Applied Physics, Eindhoven University of Technology, P.O. Box 513, 5600 MB Eindhoven, The Netherlands

[3] Hefei Innovation Research Institute, Beihang University, Hefei 230013, China

*Correspondence: Xiaoyang Lin, E-mail: XYLin@buaa.edu.cn

**ABSTRACT:**

Magnetic racetrack memory has significantly evolved and developed since its first experimental verification and is considered as one of the most promising candidates for future high-density on-chip solid state memory. However, the lack of a fast and precise magnetic domain wall (DW) shifting mechanism and the required extremely high DW motion (DWM) driving current both make the racetrack difficult to commercialize. Here, we propose a method for coherent DWM that is free from above issues, which is driven by chirality switching (CS) and an ultralow spin-orbit-torque (SOT) current. The CS, as the driving force of DWM, is achieved by the sign change of Dzyaloshinskii–Moriya interaction which is further induced by a ferroelectric switching voltage. The SOT is used to break the symmetry when the magnetic moment is rotated to the Bloch direction. We numerically investigate the underlying principle and the effect of key parameters on the DWM through micromagnetic simulations. Under the CS mechanism, a fast (~$10^2$ m/s), ultralow-energy (~5 attoJoule), and precisely discretized DWM can be achieved. Considering that skyrmions with topological protection and smaller size are also promising for future racetrack, we similarly evaluate the feasibility of applying such a CS mechanism to a skyrmion. However, we find that the CS only causes it to "breathe" instead of moving. Our results demonstrate that the CS strategy is suitable for future DW racetrack memory with ultralow power consumption and discretized DWM.

# INTRODUCTION

Magnetic racetrack memory, in which a train of up and down magnetic bits is moved electrically along a magnetic track, has become the research focus of spintronic community since its first experimental demonstration[1]. Due to its non-volatility, high storage capacity, fast speed and flexible design, it has huge application potential at various levels in the on-chip memory hierarchy[2,3]. Over the past decades, the racetrack memory has evolved[4] rapidly in terms of density[5], speed[6] and write energy[7] by employing more efficient spin torques or novel film structures. Regardless of the application direction, however, there are still several challenges before domain wall (DW) devices become commercially available. Firstly, the DW motion (DWM) needs to be highly controllable. The fast and accurate shift operations appear to be the biggest challenge and performance bottleneck from the architectural perspective[2]. Under the existing mechanism of current-driven (spin transfer torque (STT) current or spin orbit torque (SOT) current) DWM, the precise control of the DW displacement imposes further engineering challenges on the circuit/architecture level, in particular, the timing and synchronization of the current pulses or other position error correcting methods[8]. Without these measures, the typical error rate due to DW misalignment in racetrack memory will be in the range of $10^{-4}$-$10^{-5}$, far from satisfying the reliability requirements[2,9,10]. To solve this issue, researchers have proposed various methods such as creating pinning sites to control DW movement[3]. Nevertheless, the ways they manipulate the shape of the nanowires[11,12] or change their magnetic properties[13-15] often requires more complex lithography and etching processes. This also introduces additional defects and is not suitable for large-scale reproduction. Furthermore, the introduction of pinning sites will increase the energy dissipation as the DW needs higher depinning current, which further jeopardize the efficiency of DWM. Given this, other strategies to generate the accurate DWM need to be explored. Secondly, the operating power consumption of DW devices needs to be reduced. A low STT or SOT threshold current driving DWM is the key to low power DW-based devices[3,16,17]. At present, the typical value of STT[18] and SOT[19] current density for driving DWs are still in the order of $10^{11}$ A/m$^2$, which will potentially impose Joule heating and reliability issues. Different methods have been employed to reduce the current required for fast DWM, such as using a combination of anisotropy modulations[20] and spin currents. However, the efficiency gained using these methods is still not significant. Magnetic skyrmions hold promise as alternative information carriers in racetrack design due to their topologically protected stability, nanoscale size, and low drive current[21]. Under the condition of high speed movement, it has been demonstrated that the current gap for skyrmions and DWs is small[22] and the accurate shifting and the additional skyrmion Hall effect[23] cannot be solved.

Given all this, different methods need to be explored for shifting-accurate and efficient DWM strategies. The Dzyaloshinskii–Moriya interaction (DMI), as one of the origins for chiral magnetism, is currently attracting huge attention in the research community focusing on DW devices or skyrmions[24,25]. By modulating the sign and size of the DMI (characterized by the parameter $D$ (J/m$^2$)), it is possible to achieve low-power and controllable chiral DW devices[26]. Until now, several feasible methods of

DMI control, i.e., chirality switching (CS), have been reported, such as ferroelectric (FE) proximity effect[27-29], gate bias voltage control[30,31], mechanical strain[32] and hydrogen chemisorption[33]. Interesting simulation results based on CS such as driving domain walls[26] and skyrmions-based logic family[34] are also proposed. However, their proposed way of combining the FE polarization voltage and an asymmetric DMI component introduced by another bias voltage can only make the up-to-down and down-to-up DWs move in opposite directions[26]. This incoherent movement will make the mechanism of CS driven DWM unsuitable for racetrack memory, since the magnetic domain would expand or contract accordingly.

To address this, in this work we propose an ultralow power and shifting-discretized magnetic racetrack memory device via simulations, which is driven by CS and SOT. We use a very low amplitude of SOT current and a voltage pulse to enable coherent fast DWM. Different from the mainstream spin-current driving mechanism, the driving force of DWM in our device is the CS triggered by FE polarization switching[27], while the small SOT is used to break the symmetry during the CS process. First, we establish an all-oxide trilayer model compatible with existing experimental results[27,35] and demonstrate the feasibility of using it as a racetrack device. Then we explain the driving mechanism of DW under CS and the difference of symmetry breaking by SOT or magnetic field. The direction of DWM depends on the timing sequence of SOT pulse and FE voltage pulse. We explore the DW displacement under different initial and final DMI values, perpendicular magnetic anisotropy (PMA), saturation magnetization ($M_s$), and damping constant. The dependence of displacement with time proves the discrete nature of CS-driven DWM. Furthermore, a comparison of various DWM methods using the same model parameters is made to quantify the performance of our CS mechanism. In the end, we also apply the CS mechanism to a skyrmion and demonstrate its breathing behavior which differs from the DWs. These results show that our proposed device concept has the potential to empower a fast, energy efficient and shifting-accurate racetrack memory.

## MODEL AND SIMULATION

The configuration of our all-oxide trilayer device is illustrated in Figure 1a. From top to bottom, the structure is composed of $BaTiO_3$(BTO)/$SrRuO_3$(SRO)/$SrIrO_3$(SIO). The thickness of the three layers is 8, 2.4, and 8 nanometers respectively, which are taken identically from available experimental works[27,35]. The top and bottom electrodes are used to apply a FE switching voltage to the BTO and electric current to the SIO. Figure 1b shows the schematic diagram of the FE proximity effect at the BTO/SRO interface. The FE-driven ionic displacement in BTO will penetrate into SRO, which can manipulate the sign of the displacement $\delta_{Ru-O}$ between Ru and O atoms. The Hamiltonian of the DMI is defined as $H_{DMI} = \boldsymbol{D}_{12} \cdot (\boldsymbol{S}_1 \times \boldsymbol{S}_2)$, where $\boldsymbol{S}_1$ and $\boldsymbol{S}_2$ are two neighboring spins and the magnitude and sign of $\boldsymbol{D}_{12}$ are determined by the spin-orbit coupling between ferromagnetic (FM) atoms and surrounding atoms as given by $\boldsymbol{D}_{12} = D\overrightarrow{r_{12}} \times \vec{z}$. Here $\vec{z}$ represents a unit vector along the [001] axis and $\overrightarrow{r_{12}}$ points from $\boldsymbol{S}_1$ to $\boldsymbol{S}_2$, respectively[36]. Accordingly, -$\delta_{Ru-O}$ (+$\delta_{Ru-O}$) will change the sign of the vector $\vec{z}$, which via $\boldsymbol{D}_{12}$ will switch the preferred chirality of the DW eventually. The heterostructures can be epitaxially grown on a [001]-oriented $SrTiO_3$ substrate by pulsed laser deposition to ensure [001]-oriented FE polarization direction and strong PMA of SRO[27,35,37,38]. Considering the Curie temperature $T_C$ of SRO (150 K) and the large damping constant (0.43)[39] reducing the DWM distance (as the following study will prove), we choose $La_{0.67}Sr_{0.33}MnO_3$ (LSMO) as an alternative FM material. The low damping constant (0.0008-0.015)[40,41], high $T_C$ (360 K), and good structure availability with SIO[42-44] all indicate that LSMO is a suitable candidate material. It should be noted that FE proximity effect is only one of the ways to switch the DW chirality. However, all the CS ways mentioned above can be used to combine with the SOT in our model to realize deterministic and coherent DWM.

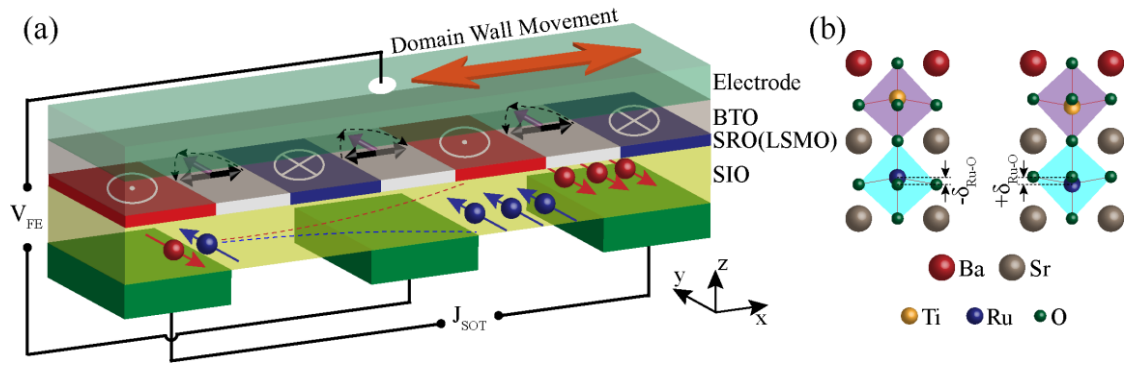

Figure 1. (a) The device structure for all-oxide trilayer DWM model driven by FE polarization switching and SOT: (from top to bottom) top electrode; FE layer $BaTiO_3$ (BTO); ferromagnetic layer $SrRuO_3$ (SRO) or $La_{0.67}Sr_{0.33}MnO_3$ (LSMO) with PMA; spin Hall layer $SrIrO_3$ (SIO); bottom electrode layer. The chiral DWs are formed due to strong DMI in the ultrathin SRO layer and the sign of $D$ can be switched by the FE polarization proximity effect of BTO. The spin current generated by SIO is used to break the symmetry along the y direction during the CS process of the DW. By the combination of the above two effects, the DWs then move in a consistent manner. The FE voltage can be applied to the whole racetrack through the SIO layer. (b) Schematic diagram of the FE proximity effect at the BTO/SRO interface. Through FE polarization, the distance of Ru and O atoms $\delta_{Ru-O}$ of the top 3-4 unit cells in the SRO will be affected by the ionic displacement in the BTO layer[27]. The positive or negative $\delta_{Ru-O}$ will ultimately determine the chirality of the DW.

The micromagnetic simulations in this work are performed using MuMax3[45] software by numerically solving the Landau-Lifshitz-Gilbert (LLG) equation with SOT and DMI terms[46] (see Supporting Information, S1). The relevant parameters in the model are listed in Table 1, which are typical values of experimentally characterized magnetic parameters. To introduce the CS process of DW, we map the polarization intensity change in the FE switching process of BTO to the DMI value change in the SRO in MuMax3. The FE switching of BTO is solved using conventional Landau-Khalatnikov (LK) equation[47,48], which is detailed in Supporting Information, S2.

Table 1. Parameters of Simulation

| Symbol | Description | Default value |
|---|---|---|
|  | FM layer area | 1024 nm × 64 nm |
| $M_S$ | Saturation magnetization[35] | $1.6 \times 10^5$ A/m |
| $A_{ex}$ | Exchange constant[27] | $2.6 \times 10^{-12}$ J/m |
| $\alpha$ | Damping constant[41] | 0.01 |
| $K$ | Magnetic anisotropy[27,29,44] | $5.7 \times 10^5$ J/m$^3$ |
| $D$ | DMI value[27] | 1.2 mJ/m$^2$ |
| $\theta_{SOT}$ | Spin Hall angle[35] | 0.58 |
| $\sigma$ | Spin polarization direction | (0 1 0) or (0 -1 0) |

To verify the feasibility of our conceptual device, the combination of FE switching voltage and SOT current are applied to drive the DW. In accordance with existing conventions, we define DW with left-handed chirality (LHC) when D<0 and right-handed chirality (RHC) when D>0. The coordinate system used here is the same as in Figure 1a. As shown in Figure 2a, excited by a FE voltage with 1 ns and a SOT current of $6 \times 10^7$ A/m$^2$, the DW changes from RHC at 0.5 ns to LHC at 2.5 ns after undergoing an intermediate state with no specific chirality and regular straight shape. Interestingly, the DW shifts 172 nm to the right during the CS process and spontaneously stops despite of presence of the voltage pulse and spin current (see supplementary video for a full DWM process). This proves that our CS-driven DWM works and the specific driving mechanism will be explained in detail later. Furthermore, we construct a nanostrip with multiple DWs and demonstrate that all chiral DWs (up-down and down-up) show unidirectional and equidistant motion. This is different from the previously reported DWM using FE-voltage to break the symmetry[26]. After having demonstrated the basic mechanism, we explore the DW displacement as a function of either a magnetic field H along y direction or an SOT current $J_{SOT}$ along x direction, as shown in Figure 2b and 2d, respectively. The results in these two figures show that DWs can move deterministically in response to CS in combination with applying H ($J_{SOT}$). The direction of DWM can be switched by changing both the directions of CS or H ($J_{SOT}$). Our DWM mechanism displays certain similarities with SOT driven DWM, in which both current direction and DW chirality determine the movement direction. In order to further demonstrate the mechanism of CS driven DWM, we performed a comparative simulation (see Figure S2) to that shown in Figure. 2a, which demonstrates that there is no contribution of SOT as driving force in our DWM. Above a certain excitation value ($\pm 0.2$ mT or $\pm 3.5 \times 10^7$ A/m$^2$), the distance of the movement tends to stabilize and the spin current breaking symmetry can move the DW farther than the magnetic field breaking symmetry. The fluctuation of the displacement with the change of H or $J_{SOT}$ should be the solution error of the MuMax3 micromagnetic solver. Interestingly, one feature should be mentioned that a magnetic field along +y and a spin current along +y ($J_{SOT}$ along +x) will cause the domain walls to move in opposite directions, which is contrary to our intuition and will be further investigated in the following.

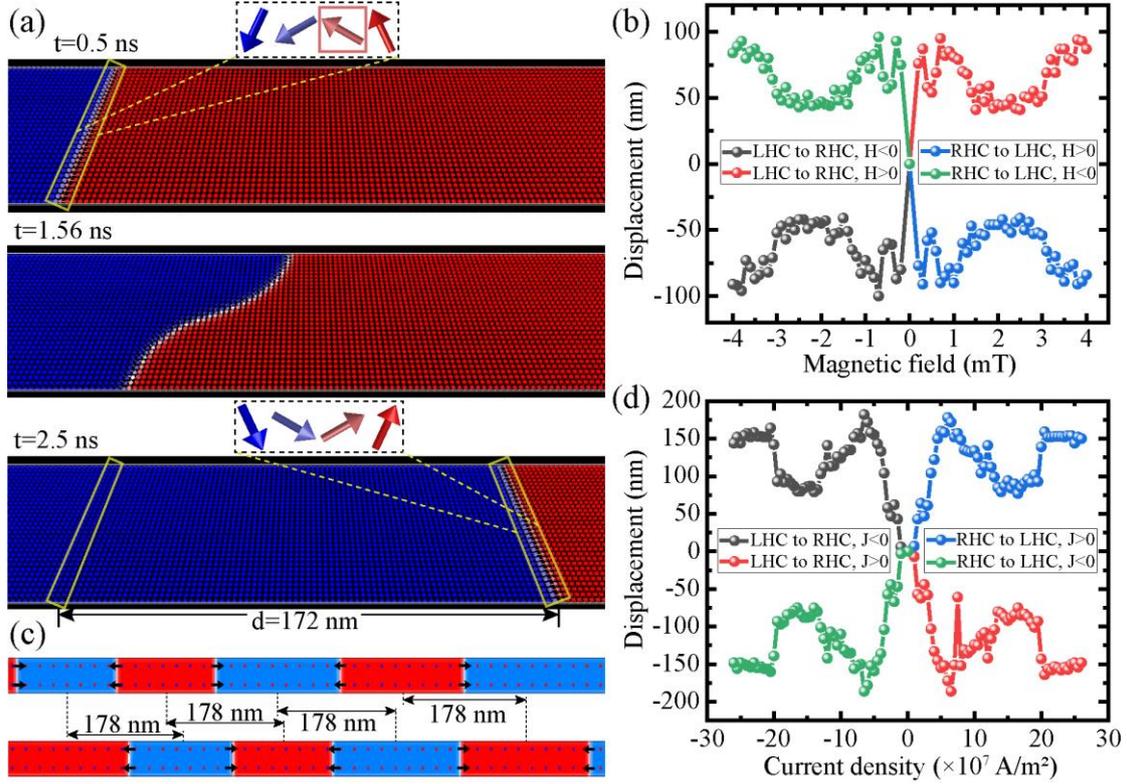

Figure 2. (a) The motion process of a DW excited by a FE voltage and a SOT current: (from top to bottom) initial state with right-handed chirality (RHC) at 0.5 ns, D = 1.2 mJ/m$^2$; intermediate state with no specific chirality and regular straight shape at 1.56 ns; final state with left-handed chirality (LHC) at 2.5 ns, D=-1.2 mJ/m$^2$. The SOT current density applied in this process is $6\times10^7$ A/m$^2$ and the CS process is completed within 1 ns. The DW eventually moves 172 nm and stop automatically. (c) Movement of a multi-domain nanometer strip driven by the CS mechanism. The result shows that the chiral DWs will move in the same direction consistently, which proves its feasibility as a racetrack memory device. (b,d) Simulated DW displacement as a function of (b) magnetic field H along y direction or (d) SOT current density J$_{SOT}$ along x direction.

## RESULTS AND DISCUSSION

Conceptually, the DWM process can be perceived as the coherent rotation of the spin magnetic moment at the center of the Neel domain[49]. To get the out-of-plane domains to move, the effective fields or torques associated working on the DW must be aligned along the domain magnetization (either up or down). To give a simplified picture of the CS driven DWM, we extract the magnetic moment at the DW position to study its dynamics. We choose the magnetic moment framed in red in Figure 2a as the object of study. Figure 3a shows the schematic of the magnetization dynamics at the

DW and Figure 3c shows the corresponding simulated moment variation with time. The magnetic dynamics can be further generalized to three processes. Firstly, as the value of $D$ shrinks from -1.2 mJ/m$^2$ to 0, the effective field of DMI along the -x direction decreases and the magnetic moment will slightly tilt upward (green arrow) in the x-z plane as DMI loses its balance against PMA. Secondly, as $D$ gradually crosses 0, the Neel morphology of the DW cannot be maintained due to energy instability and the magnetic moment rotate in the x-y plane to form a Bloch wall[50] (purple arrow). However, whether the DW turns to +y or -y direction is completely random. At this point, a weak $H_y$ or small $J_{SOT}$ is used to steer the DWs all into the same Bloch direction to break the symmetry. Lastly, as $D$ increases from 0 to 1.2 mJ/m$^2$, the magnetic moment is subjected to an increasing effective field $H_{DMI}$ along +x, which will rotate it to -z (blue arrow) due to the applied torque ($\boldsymbol{\tau} = -\boldsymbol{m_y} \times \boldsymbol{H_{DMI}}$). Figure 3c clearly shows these three processes and a magnified inset is used to show process 2. Accordingly, the DW will move constantly along +x direction until the CS process is finished. Next we focus on the odd symmetry breaking phenomenon: spin polarization along +y direction tends to rotate the DWs all to –y. Figure 3b shows the variation of magnetic component $m_y$ of the strip with time under different $J_{SOT}$ or $H_y$. For $H_y$=0 mT or $J_{SOT}$ = 3.5×10$^7$ A/m$^2$ which is below the critical value, $m_y$ will always stay at or oscillate around 0. In contrast, higher $H_y$ or $J_{SOT}$ will cause the magnetic moment to have an obvious $m_y$ component as $D$ crosses 0, whose signs are reversed.

Furthermore, we observe the direction and distance of DWM by varying the initial time of the SOT pulse as shown in Figure 3d. As the pulse initiation time goes on, the

DWM will go through three stages: stable positive displacement (0-1.4 ns), stable negative displacement (1.4-1.5 ns) and no displacement (after 1.5 ns). By comparing the variation of DW magnetic moments with the initial time of SOT pulses at 0 ns and 1.44 ns, we find that the competition between $H_{DMI}$ generated torque and SOT generated anti-damping like torque results in different motion directions (see Figure S3). At first, the magnetic moments of DW point in the -x direction due to the $H_{DMI}$:

$$H_{DMI} = \frac{D}{\mu_0 M_S \Delta} \quad (1)$$

where $\Delta = \sqrt{\frac{A_{ex}}{K_{eff}}}$ is the domain wall width and $K_{eff}$ is the effective anisotropy of the FM layer. When the $J_{SOT}$ is present from 0 ns, the DW spin will generate a magnetic moment component in the -z direction due to its effective field:

$$H_{SOT} = \frac{J_{SOT} \theta_{SOT} \hbar}{2e t_F \mu_0 M_S}(\boldsymbol{m} \times \boldsymbol{\sigma}) \quad (2)$$

where $\hbar$ is the reduced Planck constant, $e$ is the elementary charge, $t_F$ is the thickness of the FM layer, $\mu_0$ is the vacuum permeability. For $J_{SOT}=6\times10^7$ A/m², $m_z$ can reach $-2\times10^{-4}$ as shown in Figure S3d. In this case, two competing torques will exert on the DW magnetic moment, which are along –y and +y directions respectively:

$$\tau_{DMI} = -\gamma\mu_0(m_z \times H_{DMI}) \quad (3)$$

$$\tau_{SOT} = -\gamma\mu_0(m_x \times H_{SOT}) \quad (4)$$

Initially, $\tau_{DMI}$ is an order of magnitude larger than the $\tau_{SOT}$. With the CS process, $\tau_{SOT}$ gradually exceeds $\tau_{DMI}$, which changes the direction of domain wall motion. When $J_{SOT}$ is applied after $D$ switches (t=1.5 ns), the DW will not move. For $J_{SOT}=6\times10^7$ A/m² and $J_{SOT}=6\times10^8$ A/m², the critical time of transition from positive displacement to negative displacement is 1.4 ns and 1.38 ns respectively. Increasing $J_{SOT}$, i.e., increasing $\tau_{SOT}$ will advance the critical transition time, further verifying our explanations.

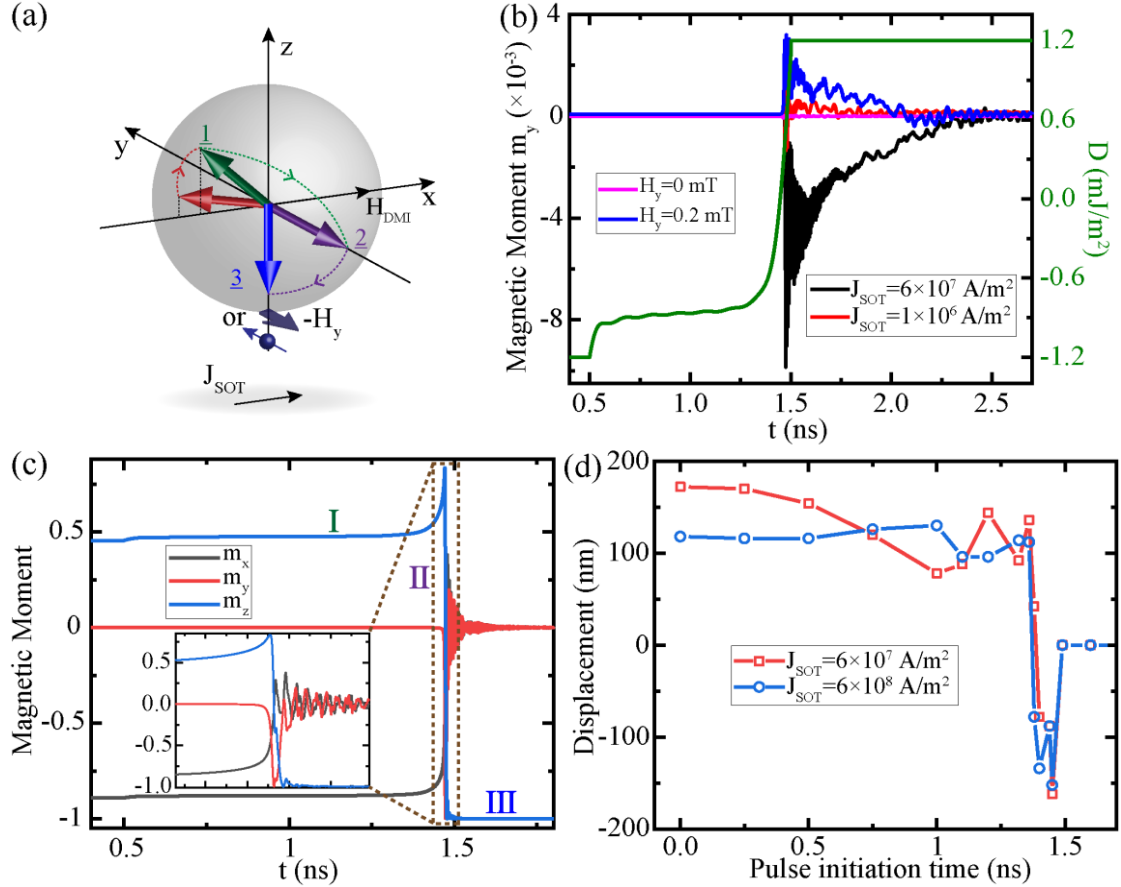

Figure 3. (a) Schematic of the magnetization dynamics at DW induced by the CS voltage and spin current (or antiparallel $H_y$). The magnetic moment goes through three processes: upward tilt (green arrow), Bloch wall (purple arrow) and downward switching (blue arrow). (b) The variation of magnetic moment component in y direction with time under different $J_{SOT}$ or $H_y$ (left axis) and the variation of DMI (right axis) with time. In this process, the reason why $m_y$ is so small is because we are counting the overall magnetic moment of the whole strip (1024 nm×64 nm) and the DW only occupies a small component. (c) Simulated magnetic moment as a function of time corresponding to the three dynamic processes in (a). (d) Simulated DW displacement as a function of pulse initiation time under two different $J_{SOT}$. For $J_{SOT}=6\times10^7$ A/m$^2$ and $J_{SOT}=6\times10^8$ A/m$^2$, the critical time of transition from positive displacement to negative displacement is 1.4 ns and 1.38 ns, respectively. Further increasing the pulse initiation time, both of the displacement turn to 0 nm.

After explaining the mechanism of DWM, we further study the influence of material parameters on the distance and velocity of DWM to shed some light on the actual experimental situation. We chose the mechanism where $\tau_{DMI}$ is the dominant torque breaking the symmetry, which does not require a precise control of the SOT pulse application time. In actual experimental situations, the absolute value of *D* are often different after the reversal due to unavoidable formed oxygen vacancies[27], defects and other factors[32,51]. As a result, we simulate the DW displacement as a function of different initial *D* and final *D* values as shown in Figure 4a. We set a DW displacement of 40 nm as the critical value for its normal function as shown in the contour line of the diagram. The upper contour line indicates that the final *D* value has the highest limit. This limitation is characterized by the DW energy:

$$\sigma_{DW} = 4\sqrt{A_{ex}K_{eff}} - \pi|D| \qquad (5)$$

When $\sigma_{DW} < 0$, the DW structure will become unstable and the magnetic domain will form a spin spiral state. The upper limit of the calculated *D* is 1.55 mJ/m², which is consistent with our simulation results. Similarly, when the initial *D* exceeds this threshold, although the DW structure is not energetically most stable, its shape can be maintained. However, before *D* turns to 0, the magnetic moment will oscillate along the -x direction due to the large $H_{DMI}$. It will cause the uncertainty when the DW turns into Bloch wall and then change the direction of DWM. As the right contour line shows, the DW will move more than 100 nm in the opposite direction with some initial *D* values. When the final *D* value is relatively small, the DW will move less than 40 nm, as shown in the contour line below. This is due to the insufficient DMI torque to switch the magnetic moment from Bloch wall to -z when the DMI effective field is comparable to the Bloch field of the DW itself. The Bloch field is produced by the magnetostatic energy associated with volume charges[50] and can be represented by

$$H_{Bloch} = \frac{4\ln(2)\,t_F\mu_0 M_S}{\pi\Delta} \qquad (6)$$

Figure 4b shows the phase diagram of the minimum current needed to drive the DW to move above 40 nm as a function of uniaxial anisotropy K and saturation magnetization $M_S$. The blank space in the upper left corner also represents the case where $\sigma_{DW} < 0$ and the DW transfers to a spin spiral state. As the uniaxial anisotropy increases, $J_{SOT}$ needs to increase accordingly to generate sufficient magnetic moment component $m_z$. With the increase of $M_S$, $H_{DMI}$ will decrease according to equation (1). It is thus necessary to increase $J_{SOT}$ to generate sufficient torque. The black area in the upper right corner indicates the case where the DW does not move even when $J_{SOT}$ reaches $1 \times 10^9$ A/m². In this situation, the DMI value needs to be further increased to generate sufficient DW driving force. Figure 4c shows the DW displacement as a function of damping constant α under $J_{SOT}=4 \times 10^7$ A/m² and $J_{SOT}=-4 \times 10^7$ A/m². Squares and triangles represent the simulation data. Red and blue lines represent their fitting curves correspondingly, which can be represented by[26] $d = \frac{\pi}{2\alpha}$. The consistency between the simulation data and the fitted curves shows that the movement distance is inversely proportional to α, which is in line with the previous theoretical explanation[26]. Then the same model is used to simulate the DWM with two different driving mechanisms: CS driven and SOT driven, as shown in Figure 4d. For CS driven DWM, the movement velocity will gradually decay and eventually reach zero. The movement is discrete and the distance is tunable by α. For SOT driven DWM, the DW moves extremely slowly before the driving current reaches a threshold ($2 \times 10^{10}$ A/m² here). The movement velocity increases initially until it reaches a stable value and then becomes constant. It will not stop until the current is turned off. The maximum speed of the two mechanisms during motion is comparable.

Now, we can systematically summarize the working mechanism of the racetrack device under our CS mechanism. Taking a material with a damping constant of 0.1 as an example, we construct a series of 16-nm-long memory bits with magnetization up or down. Each time $V_{FE}$ and $J_{SOT}$ are applied, the racetrack will perform a 1-bit shift operation precisely. If the DW needs to be continuously moved in the same direction, it is necessary to change the CS direction and $J_{SOT}$ direction at the same time. If the DW needs to be moved in an opposite direction, we change the CS direction only and keep $J_{SOT}$ the same.

It is worth noting that none of the above simulations considered the effect of temperature on the device. For a real CS-driven DW device, the thermal disturbance at room temperature may have an impact on the Bloch symmetry breaking by $J_{SOT}$. The corresponding result may be that the symmetry-breaking needs a larger $J_{SOT}$ or the actual moving distance of the device is reduced.

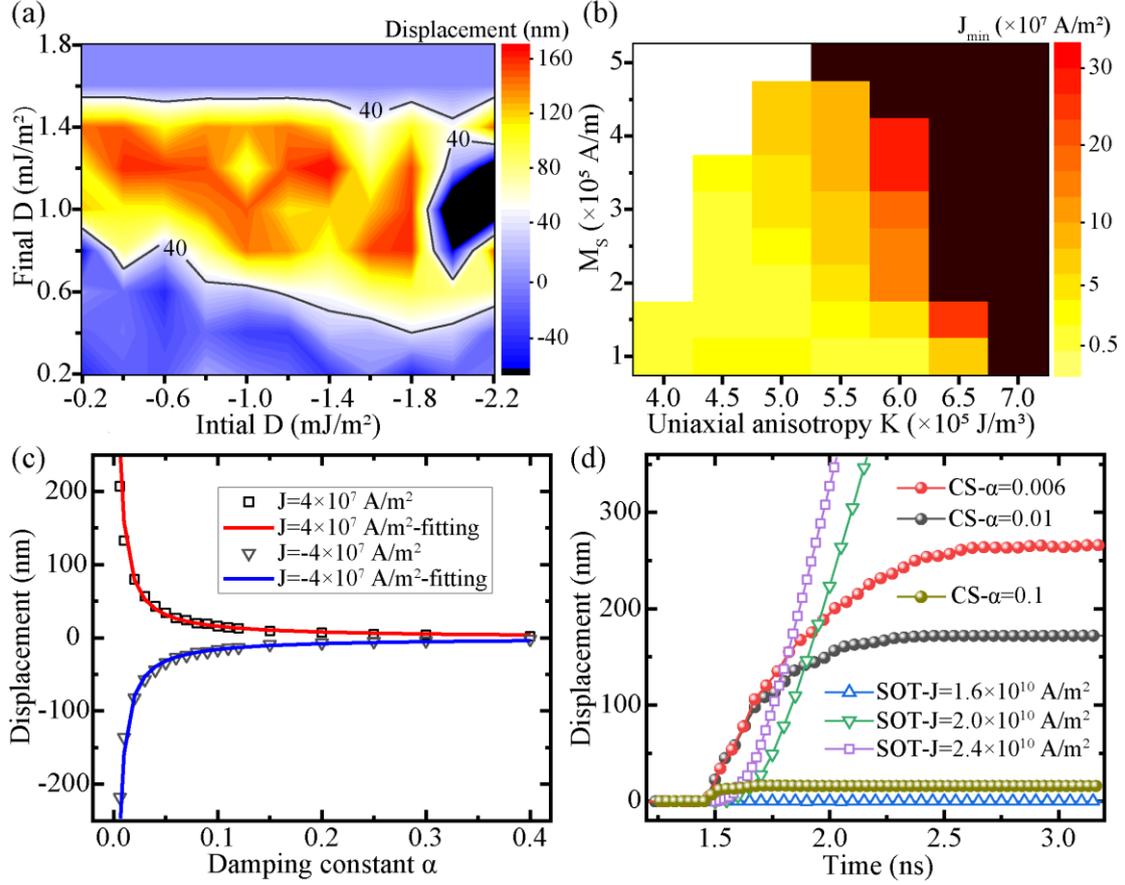

Figure 4. (a) DW displacement as a function of initial *D* and final *D* values. (b) The minimum current $J_{min}$ for driving the DW movement above 40 nm as a function of uniaxial anisotropy K and saturation magnetization $M_S$. (c) DW displacement as a function of damping constant α under different SOT current. Squares and triangles represent the simulation data. Red and blue lines represent their fitting curves correspondingly. (d) DW displacement variation with time under two different driven mechanisms: CS driven with different α ($J_{SOT}=6\times10^7$ A/m$^2$) and SOT driven with different current density.

To further evaluate the application potential of our CS mechanism, a performance comparison of STT, SOT and CS driven DWM is made using the same trilayer model and parameter settings. To make the contrast more intuitive, the movement distance is set to 172 nm (the full distance of one DW-shifting-operation driven by CS) for all mechanisms. As shown in Table 2, the results demonstrate that the current density and power consumption of CS are best among the three mechanisms, which are at least three and two orders of magnitude smaller than the existing DW driving mechanisms respectively. The average velocity of DWM driven by CS is moderate among the three mechanisms (see supporting information S5 for all the detailed energy and speed calculation process). However, the CS driven DWM is discrete, which does not require precise pulse width or creating pinning sites to force it to stop. Given all this, the DW movement based on CS has the potential to achieve ultralow power and shifting-accurate racetrack memory.

Table 2. Performance comparison of three DW driven mechanisms

| DW driven mechanism | Current density (A/m$^2$) | Speed (m/s) | Power consumption (J) | Shifting type |
| --- | --- | --- | --- | --- |
| STT | $3 \times 10^{12}$ | 143.3 | $3.982 \times 10^{-12}$ | continuous |
| SOT | $2 \times 10^{10}$ | 344 | $4.51 \times 10^{-16}$ | continuous |
| CS | $6 \times 10^{7}$ | 173.2 | $4.247 \times 10^{-18}$ | discrete |

After completing the study of driving DWM by CS, we further investigate such a CS mechanism in the case of skyrmions. Magnetic skyrmions hold promise as information carriers in racetrack memory due to their advantages in topologically protected stability, nanoscale size and low driving current compared to DWs[21]. However, our simulation results show that skyrmions are not suitable to be driven by our CS mechanism. Figure 5(a), (b) and (c) show the evolution process of a skyrmion excited by a FE switching voltage only. The skyrmion will undergo a breath and a transition from Neel to Bloch and back to Neel. Its initial diameter and maximum diameter during breathing are 12 nm and 126 nm respectively. This is completely different from the behavior of magnetic DWs. We attribute this "breath" to the fact that $H_y$ or $J_{SOT}$ fails to break the symmetry along y direction. When a Neel skyrmion is gradually transformed into a Bloch skyrmion, it will spontaneously form a clockwise or counterclockwise arrangement due to its own topological protection, which will cause the magnetic moments at any two ends to always point into the antiparallel direction (for example the magnetic moments at +x and –x will point to +y and –y directions respectively). The result is that the skyrmion doesn't need a symmetry breaking magnetic field or current and the up-to-down and down-to-up DWs always move in opposite directions. In addition, we plot the initial diameter $d_{int}$ and the maximum diameter $d_{max}$ during breathing of the skyrmion as a function of the DMI value. The $d_{int}$ increases with the value of $D$ as demonstrated in previous articles[52]. The $d_{max}$ will also increase with the value of $D$ because the energy gap produced by the CS increases. The inset shows some schematic diagrams of intermediate and final states of skyrmions. For the above reasons, skyrmions hardly move whether or not a symmetry-breaking magnetic field or current is applied. Figure 5(e) shows the initial and final state of a skyrmion excited by a FE switching voltage and a magnetic field $H_y$ of 1.2 T. Even if $H_y$ or $J_{SOT}$ is large enough (1.2 T or $1 \times 10^{10}$ A/m$^2$) to deform the skyrmion, it will not move more than 5 nm during the CS process.

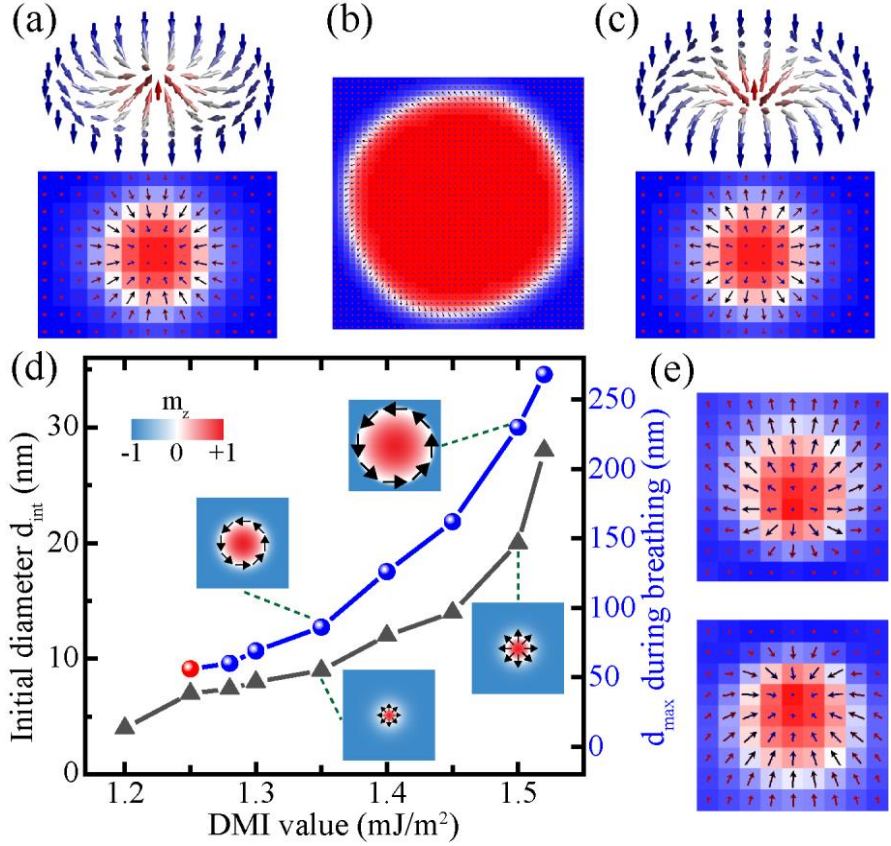

Figure 5. CS process of magnetic skyrmions. (a,b,c) The evolution process of skyrmion excited by a FE switching voltage only: initial state with LHC at 1 ns, D = -1.4 mJ/m$^2$ (a); intermediate state which is a gradually expanding Bloch skyrmion at 1.2 ns (b); final state with RHC at 2.5 ns, D = 1.4 mJ/m$^2$ (c). The upper and lower parts of (a) and (c) represent the schematic diagram and the actual simulation figure of the skyrmion, respectively. During the entire process, the skyrmion will take one breath, expanding its diameter from 12 nm to a maximum of 126 nm and then shrinking to 12 nm. Its type will go through a transition from Neel to Bloch and then to Neel again. The skyrmion will not move in this whole process. (d) Initial diameter $d_{int}$ and maximum diameter $d_{max}$ during breathing of the skyrmion as a function of DMI values. The blue (red) spheres and black triangles represent $d_{max}$ and $d_{int}$ respectively. For D = 1.2 mJ/m$^2$, the skyrmion will annihilate before expanding. For D = 1.25 mJ/m$^2$, the skyrmion will annihilate after expanding to 56 nm. All blue spheres indicate that skyrmions do not annihilate and return to their original diameter after the CS process. Inset: schematic diagrams of intermediate and final states of skyrmions at some typical D values. (e) Initial and final state of a skyrmion excited by a FE switching voltage and a magnetic field $H_y$ of 1.2 T. Even if the $H_y$ or $J_{SOT}$ is increased (1.2 T or 1×10$^{10}$ A/m$^2$) to deform the skyrmion, it will hardly move during the CS process.

## CONCLUSIONS

To make a conclusion, in view of reducing the power consumption and imposing a fast and precise shifting of domains in racetrack memory, we propose a mechanism of DWM driven by CS and low-amplitude symmetry-breaking SOT current. The device model is established and the feasibility of the device as a racetrack memory is demonstrated. We explore the intrinsic mechanism of CS-driven DWM and simulate the effect of key parameters on device performance. Compared to conventional DW driven mechanisms (STT or SOT), the advantages of this method is demonstrated through simulation in which fast (~170 m/s), ultralow-energy (~5 attoJoule per 172 nm step-shifting), and discrete DWM can be achieved. Our CS-driven mechanism can pave a promising way for future ultralow power and shifting-accurate racetrack memories.

## SUPPORTING INFORMATION

LLG equations with SOT and DMI terms; the FE switching of BTO using LK equation; comparative simulation results to Figure. 2a; DW magnetic moments variation with the initial time of SOT pulses at 0 ns and 1.44 ns; the specific calculation process of the performance of the three DW driving mechanisms.

## ACKNOWLEDGMENTS

This work was supported in part by the Beijing Natural Science Foundation (No. 4232070), National Natural Science Foundation of China (No. 51602013 and 52261145694), the International Mobility Project (No. B16001), the China Scholarship Council (CSC) and the European Union's Horizon 2020 research and innovation programme under the Marie Sklodowska-Curie grant agreement No. 860060.

## AUTHOR INFORMATION

**Corresponding Authors**

**Xiaoyang Lin** - School of Integrated Circuit Science and Engineering, Beihang University, Beijing, 100191, China; Hefei Innovation Research Institute, Beihang University, Hefei 230013, China; orcid.org/ 0000-0002-2062-0050; E-mail: XYLin@buaa.edu.cn

**Authors**

**Shen Li** - School of Integrated Circuit Science and Engineering, Beihang University, Beijing, 100191, China; Department of Applied Physics, Eindhoven University of Technology, P.O. Box 513, 5600 MB Eindhoven, The Netherlands; Hefei Innovation Research Institute, Beihang University, Hefei 230013, China; orcid.org/ 0000-0002-9482-3045

**Pingzhi Li** - Department of Applied Physics, Eindhoven University of Technology, P.O. Box 513, 5600 MB Eindhoven, The Netherlands

**Suteng Zhao -** School of Integrated Circuit Science and Engineering, Beihang University, Beijing, 100191, China

**Zhizhong Si** - School of Integrated Circuit Science and Engineering, Beihang University, Beijing, 100191, China

**Guodong Wei** – School of Integrated Circuit Science and Engineering, Beihang University, Beijing, 100191, China

**Bert Koopmans** - Department of Applied Physics, Eindhoven University of Technology, P.O. Box 513, 5600 MB Eindhoven, The Netherlands

**Reinoud Lavrijsen** - Department of Applied Physics, Eindhoven University of Technology, P.O. Box 513, 5600 MB Eindhoven, The Netherlands

**Weisheng Zhao** - School of Integrated Circuit Science and Engineering, Beihang University, Beijing, 100191, China; Hefei Innovation Research Institute, Beihang University, Hefei 230013, China; orcid.org/0000-0001-8088-0404;

**Conflict of interest**

The authors declare that they have no conflict of interest.

844.